# Spectroscopic confirmation of linear relation between Heisenberg- and interfacial Dzyaloshinskii-Moriya-exchange in polycrystalline metal films

Hans T. Nembach, Justin M. Shaw, Mathias Weiler, Emilie Jué and Thomas J. Silva

*Electromagnetics Division, National Institute of Standards and Technology,*
*Boulder, Colorado 80305, USA*

Proposals for novel spin-orbitronic logic[1] and memory devices[2] are often predicated on assumptions as to how materials with large spin-orbit coupling interact with ferromagnets when in contact. Such interactions give rise to a host of novel phenomena, such as spin-orbit torques[3,4], chiral spin-structures[5,6] and chiral spin-torques[7,8]. These chiral properties are related to the anti-symmetric exchange, also referred to as the interfacial Dzyaloshinskii-Moriya interaction (DMI)[9,10]. For numerous phenomena, the relative strengths of the symmetric Heisenberg exchange and the DMI is of great importance. Here, we use optical spin-wave spectroscopy (Brillouin light scattering) to directly determine the DMI vector $\vec{D}$ for a series of $Ni_{80}Fe_{20}$/Pt samples, and then compare the nearest-neighbor DMI coupling energy with the independently measured Heisenberg exchange integral. We find that the $Ni_{80}Fe_{20}$-thickness-dependencies of both the microscopic symmetric- and antisymmetric-exchange are identical, consistent with the notion that the basic mechanisms of the DMI and Heisenberg exchange essentially share the same underlying physics, as was originally proposed by Moriya[11]. While of significant fundamental importance, this result also leads us to a deeper understanding of DMI and how it could be optimized for spin-orbitronic applications.

Recent experimental results have demonstrated how the interplay of symmetric (Heisenberg) exchange and anti-symmetric (DMI) exchange together with anisotropy can give rise to a variety of magnetostatic phenomena, such as magnetic skyrmion lattices[12], spiral spin structures[13] and chiral domain walls[14]. In bilayer materials with a sufficiently thin, perpendicular magnetized ferromagnet (FM) adjacent to a metal with large spin-orbit coupling in the conduction band, a large DMI favors Néel domain walls with a fixed chirality[15] as opposed to Bloch walls. The combination of a chiral domain wall structure and spin-orbit torque can give rise to fast current induced domain wall motion[3]. The direction and the speed are both dependent on the sign and the strength of the DMI and the spin-orbit torque[8,7]. Moreover, theory for a Rashba model predicts that the interfacial spin-orbit torque is proportional to the ratio of symmetric and anti-symmetric exchange[16]. Thus, direct determination of both the DMI and Heisenberg exchange is crucial for the understanding of the underlying physics in such materials systems and a better understanding of the spin-orbit torques.

To date, direct measurements of anti-symmetric exchange are limited to exotic measurement techniques that can only be applied to a few highly specialized sample systems. For example, the DMI constant has been measured via synchrotron-based X-ray scattering interferometry for the weak ferromagnet $FeBO_3$[17], by spin-polarized electron energy loss spectroscopy for an atomic bilayer of Fe on W(110) [18] and by spin-polarized scanning tunneling microscopy for atomic monolayer Mn on W(110)[5]. Until now estimation of the anti-symmetric exchange in the case of arbitrary materials has only been possible via inference from indirect measurement methods. These methods include both determination of the critical ferromagnetic layer thickness at the transition from a Néel to a Bloch wall[14], and measurements of domain wall motion [19,20,21]. Furthermore, quantitative experimental comparison of the symmetric and anti-symmetric exchange is still outstanding.

Recent theory predicts an asymmetric dispersion shift of long-wavelength thermal spin-waves in magnetic thin films due to the DMI[22,23]. Motivated by this theory, we used Brillouin light scattering (BLS) to directly measure the predicted asymmetric dispersion shift, which in turn allowed us to determine the magnitude and direction of the DMI vector in a technological relevant sample system: we used a series of sputtered multilayer stacks consisting of $SiN/Ni_{80}Fe_{20}(t)/Pt(6\ nm)/Ta(3\ nm)$/substrate where $t$ ranged from 1 nm to 13 nm. We independently determined the symmetric exchange by fitting of low-temperature magnetometry data to the Bloch $T^{3/2}$ law for the same samples. Comparison of the two data sets allowed us to unambiguously determine the proportionality of the symmetric and antisymmetric exchange.

The Hamiltonian for the DMI between two spins $\vec{S}_i$ and $\vec{S}_j$ on the atomic sites $i$ and $j$ in a thin ferromagnetic film adjacent to a high spin-orbit material is given by $\mathrm{H}_{DMI} = -\vec{D}_{ij} \cdot (\vec{S}_i \times \vec{S}_j)$. $\vec{D}_{ij} = D_{ij}\vec{n} \times \vec{e}_{ij}$ is the DMI vector, which lies in the symmetry breaking plane and is perpendicular to the unit vector that connects sites $i$ and $j$, $\vec{e}_{ij} \doteq \vec{r}_{ij}/|\vec{r}_{ij}|$, as shown in Figs. 1**a** and **b**. $\vec{n}$ is the interface normal. The DMI is a three-site exchange mechanism that is mediated by an atom in the high spin-orbit material. Spin-waves have a spatial chirality, which depends on their propagation direction with respect to the direction of the magnetization. In particular, for $\vec{M} \parallel +z$, spin waves propagating along -$x$ have counter-clockwise spatial chirality (Fig. 1**a**), while those propagating along +x have clockwise chirality (Fig. 1**b**). The spatial chirality favored by the DMI is fixed and thus the presence of the DMI causes an asymmetric modification of the spin-wave dispersion relation. For an in-plane magnetized film with spin-waves propagating perpendicular to the magnetization direction, the DMI modifies the frequency

$$f_M = f_0 + \Delta f_{DMI} \quad (1)$$

of the spin waves[22] (See middle panel of Fig. 1c.) Here, $f_0$ is the spin-wave frequency in the absence of the DMI, and

$$\Delta f_{DMI} = \left| \frac{g^\| \mu_B}{h} \right| \text{sgn}(M_z) \frac{D_{DMI}}{2M_s} k \quad (2)$$

is the DMI induced frequency shift, where $D_{DMI} \propto D_{ij}$ is the volumetric DMI constant that determines the sign and magnitude of the DMI vector. $g^\|$ is the in-plane spectroscopic splitting factor, $M_s$ the saturation magnetization, $\vec{k}$ the wavevector of the spin-waves, $\mu_B$ the Bohr magneton, and $h$ Planck's constant. In the presence of the DMI, the spin-wave frequencies have a shift linear in $k$ to either higher or lower frequencies, depending of the propagation direction and the polarity of the static magnetization component $\vec{M}$. If a spin-wave has the same spatial chirality as the DMI, the spin-wave frequency is reduced. Conversely, it is increased for the opposite chirality.

The dispersion characteristics of spin-waves in thin magnetic films can be measured with BLS. In our BLS measurements, a laser is focused onto a sample, and the photons are inelastically back-scattered by the quantized spin-waves, i.e. magnons. Momentum conservation dictates that magnons propagating towards the incoming laser beam must be annihilated (the anti-Stokes process), and magnons propagating in the opposite direction must be created (the Stokes process). (See Fig.1c.) Energy conservation then uniquely identifies the inelastic energy shift of the back-scattered photons with the magnon propagation direction: if the energy of the scattered photon is increased (or decreased) by the magnon energy, the direction of magnon propagation is either towards (or away from) the laser beam. Thus the measurement of the scattered photon energy can be used to determine the frequency difference of spin-waves propagating in opposite directions, as sketched in Fig. 1c. We measured the spin-wave frequencies for the two field polarities transverse to the scattering plane, and for the two propagation directions. This yields four independent measurements of spin-wave frequency as a function of magnetization polarity and propagation direction. In Fig. 2a and 2b, the normalized BLS spectra for $Ni_{80}Fe_{20}$/Pt samples with a $t$ = 1.3 nm- and $t$ = 2.0 nm-thick $Ni_{80}Fe_{20}$ are shown. For comparison, we also show data for a control sample of 2.0 nm thick Ni$_{80}$Fe$_{20}$ without a Pt underlayer in Fig.2**c**. The BLS spectra of the samples with a Pt underlayer exhibit a frequency shift in the order of 100 MHz that changes sign with magnetization polarity or spin wave propagation direction reversal. This frequency difference is larger for the sample with thinner Ni$_{80}$Fe$_{20}$, as expected for an interfacial source of symmetry breaking, whereas there is no frequency difference for the control sample, in agreement with the absence of interfacial DMI in this case.

In Fig. 3 we show the *t* dependence of the frequency shift $\Delta f_{\text{DMI}}$ averaged over the four possible combinations of the magnetization polarity and propagation direction. Consistent with the interfacial nature of the DMI, $\Delta f_{\text{DMI}}$ decreases with increasing *t*. According to Eq. (2) $M_s$ and $g^{\parallel}$ are required in order to determine $D_{\text{DMI}}$ from the measured $\Delta f_{\text{DMI}}$. We thus measured $M_s$ at 300 K via SQUID magnetometry, with the results shown in Fig. 4**a** and $g^{\parallel}$ by ferromagnetic resonance spectroscopy. We then calculate $D_{\text{DMI}}$ shown in Fig. 4**b**. Under the assumption that the DMI is a strictly interfacial property that is not affected by the bulk properties of the magnetic film, the magnitude of $D_{\text{DMI}}$ should be linearly proportional to $1/t$ and be zero in the limit of an infinitely thick film. However, the measured $D_{\text{DMI}}$ exhibits a non-linear dependence on $1/t$. This result implies that the DMI strength at the interface itself changes with the film thickness. Under the assumption that only the first monolayer of Ni$_{80}$Fe$_{20}$ at the Pt interface contributes to the antisymmetric exchange, the strength of the DMI at the interface $D_{\text{DMI}}^{\text{int}}$ can be calculated from

$$D_{\text{DMI}}^{\text{int}} = D_{\text{DMI}} \cdot \frac{\sqrt{3}}{a} t, \quad (2)$$

where $a$ = 0.354 nm is the lattice constant for Ni$_{80}$Fe$_{20}$. $D_{\text{DMI}}^{\text{int}}$ is shown in Fig. 4**c** (left scale) and has a non-trivial dependence on the Ni$_{80}$Fe$_{20}$ thickness. In particular, we find that the magnitude of the antisymmetric exchange varies by a factor of almost 2.5 over the range of sample thicknesses. By use of Eq. (S16) in the SI, the largest value for the interatomic antisymmetric exchange energy $D_{\text{nn}}$ that we obtain (for $t = 6.1$ nm) is -8 meV. This is very similar to the recently reported result in Ref. [14] of -9.2 meV for an 8.4-monolayer stack of Co/Ni on Pt (111), though it is substantially larger than the value of +0.9 meV for Fe/W in Ref. [18].

Theoretical calculations for the Dzyaloshinskii-Moriya interaction in bulk systems predict that the symmetric- and anti-symmetric-exchange are proportional to each other [11,24,25]. It is therefore reasonable to speculate that the thickness-dependence of $D_{int}$ is the result of an unexpected thickness-dependence of the symmetric exchange for this particular system. The Heisenberg exchange Hamiltonian that describes the interaction of two spins is given by $H_{ex} = -J_{ij} \vec{S}_i \cdot \vec{S}_j$, with the exchange integral $J_{ij} = 2g^{\|} \mu_B D_{spin} / (M_s a^5)$ for the fcc lattice, where $D_{spin}$ is the spin-wave stiffness.

We measure the temperature-dependence of the saturation magnetic moment $m_s(T)$ with SQUID and fit the data with the Bloch $T^{3/2}$ law at low temperatures,

$$m_s(0 \text{ K}) - m_s(T) \propto \left( \frac{k_B T}{D_{spin}^{0 \text{ K}}} \right)^{3/2}, \quad (3)$$

where $T$ is the temperature, $D_{spin}^{0 \text{ K}}$ is the low-temperature spin-wave stiffness, and $k_B$ is the Boltzmann constant[26]. In order to determine the exchange constant $A$ at 300 K, we use

$$A \doteq A(T = 300 \text{ K}) = \frac{M_s^{300 \text{ K}} D_{spin}^{300 \text{ K}}}{2 g^{\|} \mu_B}, \quad (5)$$

where we take the temperature-dependent renormalization of the spin-wave stiffness into account[27]. In Fig. 4c (right scale), the thickness-dependence of $A$ is shown. While the microscopic origins of the variation in the symmetric exchange with film thickness is unclear, it is an empirical fact that both the symmetric exchange $A$ and the anti-symmetric exchange $D_{DMI}^{int}$ exhibit the same nontrivial functional dependence on reciprocal thickness. If we consider only nearest neighbor interactions, the ratio $D_{DMI}/A$ is proportional to the ratio of the nearest-neighbor exchange integral $J_{nn}$ and the magnitude of the nearest-neighbor DMI $D_{nn}$,

$$\frac{D_{nn}}{J_{nn}} = \frac{2\sqrt{2} a}{3} \cdot \frac{D_{DMI}^{int}}{A}, \quad (6)$$

as shown in the SI. From $D_{DMI}^{int}$ and $A$ in Fig. 4c, we find by fitting Eq. (6) to the data that $D_{nn}/J_{nn} = -(30 \pm 1)\%$ independent of $Ni_{80}Fe_{20}$ thickness. The linear proportionality between the symmetric and antisymmetric exchange was first proposed for bulk materials in the original theory of Moriya[11]. Similarly, such a proportionality was also predicted by Fert *et al*[24,25], for metallic spin-glass systems. Our results confirm that this proportionality also applies to two-dimensional systems at the interface between a ferromagnetic layer and a material with large spin-orbit coupling.

An important implication of our result is the independence of interfacial chiral ordering on parameters that affect the Heisenberg exchange, as the spin canting angle at adjacent atomic sites is proportional to the ratio $D_{\text{DMI}}^{\text{int}}/A$. Thus, the methods and results presented here not only serve to elucidate the fundamental properties of the DMI, but may also have a substantial impact on the methodology to successfully develop spin-orbitronic devices.

**Contributions:** H.N. conceived the experiment, performed the BLS measurements and analyzed the BLS data. J.S. fabricated and characterized the samples and performed SQUID measurements, M.W. performed the FMR measurements and analysis. All authors contributed to the interpretation of the results and writing of the manuscript.

**Methods**

**Sample preparation**

All samples were prepared by dc magnetron sputtering in an Ar base pressure of $\approx 0.07$ Pa ($\approx 0.5$ mTorr) and a chamber base pressure of $3 \cdot 10^{-6}$ Pa ($2 \cdot 10^{-8}$ Torr). Samples were rotated at 1 Hz - 2 Hz during deposition to eliminate growth-induced anisotropy. The deposition rates were calibrated by X-ray reflectometry. The SiN(5 nm-10 nm)/Ni$_{80}$Fe$_{20}$($t$)/Pt(6 nm)/Ta(3 nm)/substrate films were deposited on thermally oxidized Si. The Ta seed layer induced a strong (111)-texture of both the Pt and Ni$_{80}$Fe$_{20}$.

**SQUID Magnetometry**

We measured in-plane hysteresis curves at room temperature to determine the saturation magnetization $M_s$(300 K) of our samples. We find that $M_s$(300 K) decreases with decreasing Ni$_{80}$Fe$_{20}$ layer thickness. The saturation magnetization $M_s$(0 K) was determined from fits of the temperature $T$ dependent magnetic moment $m_s(T)$ to the Bloch T$^{3/2}$ law. In order to determine from these fits the spin-wave stiffness $D_{spin}^{0K}$ at 0 K for the Ni$_{80}$Fe$_{20}$ thickness series, we first calculate the density of excited magnons at a given temperature. This calculation takes into account finite size effects due to the reduced thickness of the Ni$_{80}$Fe$_{20}$ layer into account. The room temperature exchange constant $A$ is then determined using a mean-field approach for the temperature dependence of the exchange. The details on the determination of the exchange constant, renormalization and calculation of the magnon density are given in the Supplemental Information.

**Brillouin Light Scattering Spectroscopy**

We use a Brillouin light scattering spectrometer with a 6-pass, tandem, Fabry-Perot interferometer to measure the thermal spin-waves frequency at a fixed angle of incidence $\theta = 45°$. For our measurements the incident laser power was 40 mW. A $\lambda$=532 nm laser beam was focused on the sample with an $f$/1.2 lens. Thus the wavevector of the measured spin-waves is $k$=16.7 µm$^{-1}$. The collimated backscattered light was spatially filtered with a 25 mm diameter aperture to reduce the wavenumber uncertainty.

**Ferromagnetic Resonance Measurements**

We employed ferromagnetic resonance spectrometers with the external magnetic field both parallel and perpendicular to the sample plane in order to determine the respective spectroscopic splitting factors $g^{\perp}$ and $g^{\parallel}$ and the perpendicular anisotropy field $H_k$, . Details about the measurements are given in the Supplemental Information.

If not otherwise stated, all error bars and all uncertainties stated in the text are single standard deviation uncertainties.

**Acknowledgement**

The authors are grateful to Ward Johnson for his support of our BLS measurements, and thank Mark Stiles and Robert McMichael for stimulating discussions.

1. Bhowmik, D., You, L. & Salahuddin, S. Spin Hall effect clocking of nanomagnetic logic without a magnetic field. *Nat. Nanotechnol.* **9,** 59–63 (2013).

2. Parkin, S. S. P., Hayashi, M. & Thomas, L. Magnetic Domain-Wall Racetrack Memory. *Science* **320,** 190–194 (2008).

3. Miron, I. M. *et al.* Perpendicular switching of a single ferromagnetic layer induced by in-plane current injection. *Nature* **476,** 189–193 (2011).

4. Liu, L. *et al.* Spin-Torque Switching with the Giant Spin Hall Effect of Tantalum. *Science* **336,** 555–558 (2012).

5. Bode, M. *et al.* Chiral magnetic order at surfaces driven by inversion asymmetry. *Nature* **447,** 190–193 (2007).

6. Chen, G. *et al.* Novel Chiral Magnetic Domain Wall Structure in Fe/Ni/Cu(001) Films. *Phys. Rev. Lett.* **110,** (2013).

7. Ryu, K.-S., Thomas, L., Yang, S.-H. & Parkin, S. Chiral spin torque at magnetic domain walls. *Nat. Nanotechnol.* **8,** 527–533 (2013).

8. Emori, S., Bauer, U., Ahn, S.-M., Martinez, E. & Beach, G. S. D. Current-driven dynamics of chiral ferromagnetic domain walls. *Nat. Mater.* **12,** 611–616 (2013).

9. Dzyaloshinsky, I. A thermodynamic theory of 'weak' ferromagnetism of antiferromagnetics. *J. Phys. Chem. Solids* **4,** 241–255 (1958).

10. Moriya, T. New Mechanism of Anisotropic Superexchange Interaction. *Phys. Rev. Lett.* **4,** 228–230 (1960).

11. Moriya, T. Anisotropic Superexchange Interaction and Weak Ferromagnetism. *Phys. Rev.* **120,** 91–98 (1960).

12. Mühlbauer, S. *et al.* Skyrmion Lattice in a Chiral Magnet. *Science* **323,** 915–919 (2009).

13. Ferriani, P. *et al.* Atomic-Scale Spin Spiral with a Unique Rotational Sense: Mn Monolayer on W(001). *Phys. Rev. Lett.* **101,** 027201 (2008).

14. Chen, G. *et al.* Tailoring the chirality of magnetic domain walls by interface engineering. *Nat. Commun.* **4,** (2013).

15. Thiaville, A., Rohart, S., Jué, É., Cros, V. & Fert, A. Dynamics of Dzyaloshinskii domain walls in ultrathin magnetic films. *EPL Europhys. Lett.* **100,** 57002 (2012).

16. Kim, K.-W., Lee, H.-W., Lee, K.-J. & Stiles, M. D. Chirality from Interfacial Spin-Orbit Coupling Effects in Magnetic Bilayers. *Phys. Rev. Lett.* **111,** (2013).

17. Dmitrienko, V. E. *et al.* Measuring the Dzyaloshinskii-Moriya interaction in a weak ferromagnet. *Nat. Phys.* **10,** 202–206 (2014).

18. Zakeri, K. *et al.* Asymmetric Spin-Wave Dispersion on Fe(110): Direct Evidence of the Dzyaloshinskii-Moriya Interaction. *Phys. Rev. Lett.* **104,** 137203 (2010).

19. Franken, J. H., Herps, M., Swagten, H. J. M. & Koopmans, B. Tunable chiral spin texture in magnetic domain-walls. *Sci. Rep.* **4,** (2014).

20. Je, S.-G. *et al.* Asymmetric magnetic domain-wall motion by the Dzyaloshinskii-Moriya interaction. *Phys. Rev. B* **88,** 214401 (2013).

21. Hrabec, A. *et al.* Measuring and tailoring the Dzyaloshinskii-Moriya interaction in perpendicularly magnetized thin films. *Phys. Rev. B* **90,** 020402 (2014).

22. Moon, J.-H. *et al.* Spin-wave propagation in the presence of interfacial Dzyaloshinskii-Moriya interaction. *Phys. Rev. B* **88,** 184404 (2013).

23. Kostylev, M. Interface boundary conditions for dynamic magnetization and spin wave dynamics in a ferromagnetic layer with the interface Dzyaloshinskii-Moriya interaction. *J. Appl. Phys.* **115,** 233902 (2014).

24. Fert, A., Cros, V. & Sampaio, J. Skyrmions on the track. *Nat. Nanotechnol.* **8,** 152–156 (2013).

25. Fert, A. Magnetic and transport Properties of metallic multilayers. *Mater. Sci. Forum* **59&60,** 439

26. Vaz, C. a. F., Bland, J. a. C. & Lauhoff, G. Magnetism in ultrathin film structures. *Rep. Prog. Phys.* **71,** 056501 (2008).

27. Atxitia, U. *et al.* Micromagnetic modeling of laser-induced magnetization dynamics using the Landau-Lifshitz-Bloch equation. *Appl. Phys. Lett.* **91,** 232507 (2007).

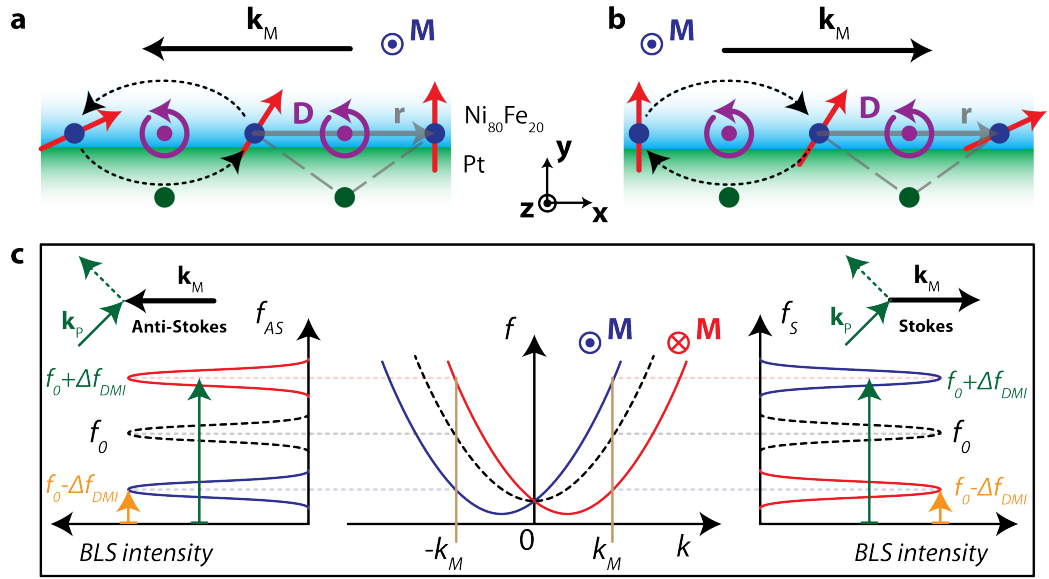

Figure 1: **Modification of spin-wave propagation in the presence of interfacial DMI**. **a** Sketch of a Damon-Eshbach spin wave propagating at the Ni$_{80}$Fe$_{20}$/Pt interface with wavevector $\vec{k}_M \parallel -x$ with the magnetization $\vec{M} \parallel +z$. The canted arrows depict the dynamic components of the spins at a snapshot in time. The dashed arrows indicate the spatial chirality of the spin wave along $x$. Any two neighboring spins at FM sites (blue atoms), are coupled by a DMI vector $D$ (purple vector pointing out of page) via a three-site exchange mechanism that includes a Pt atom (green atoms). The Pt atom serves to both break the symmetry and provide the necessary spin-orbit coupling. The prefered chirality of the antisymmetric exchange indicated by the purple arrow circulating about the DMI vector is identical to the spin wave chirality. **b** For $\vec{k}_M \parallel +x$, the spatial chirality of the spin wave is opposite to that favored by the DMI. **c** The central panel shows schematic spin-wave dispersion curves in the absence of DMI (dashed) and with DMI (solid) for $\vec{M} \parallel \pm z$, respectively. A sketch of the expected BLS spectra for $-\vec{k}_M$ (Anti-Stokes process, annihilation of a magnon) and $+\vec{k}_M$ (Stokes process, generation of a magnon) is shown on the left and right side of the lower panel, respectively

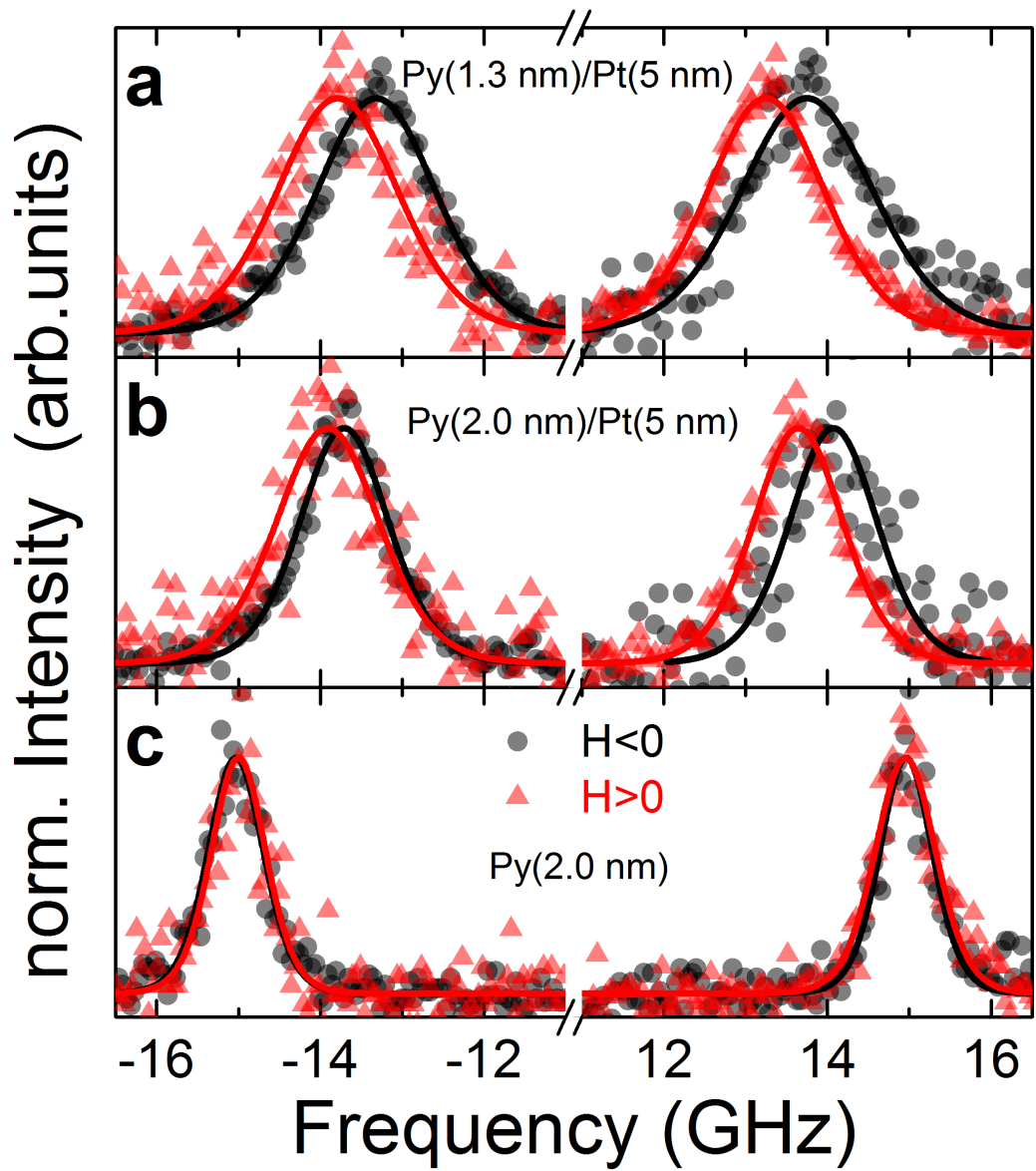

Figure 2: **Normalized data of spin-wave spectra, as measured by BLS.** The measurements were carried out for the two opposite magnetization polarities (circles and triangles). The lines are fits to the experimental data. The peak positions correspond to the frequencies of Stokes (negative frequencies) and anti-Stokes (positive frequencies) processes with a fixed wavevector of $|k|=16.7$ μm$^{-1}$. **a** Data for a 1.3 nm-thick Ni$_{80}$Fe$_{20}$ film with a 5 nm-thick Pt underlayer. **b** Data for a 2.0 nm Ni$_{80}$Fe$_{20}$ film with a 5 nm Pt underlayer. The spin-wave frequency is clearly shifted by reversal of the magnetization direction in both **a** and **b**, and the frequency shift is reduced for the sample with 2.0 nm of Ni$_{80}$Fe$_{20}$. **c** Data for a reference sample without Pt do not show a frequency shift that depends on the magnetization direction or spin-wave propagation direction.

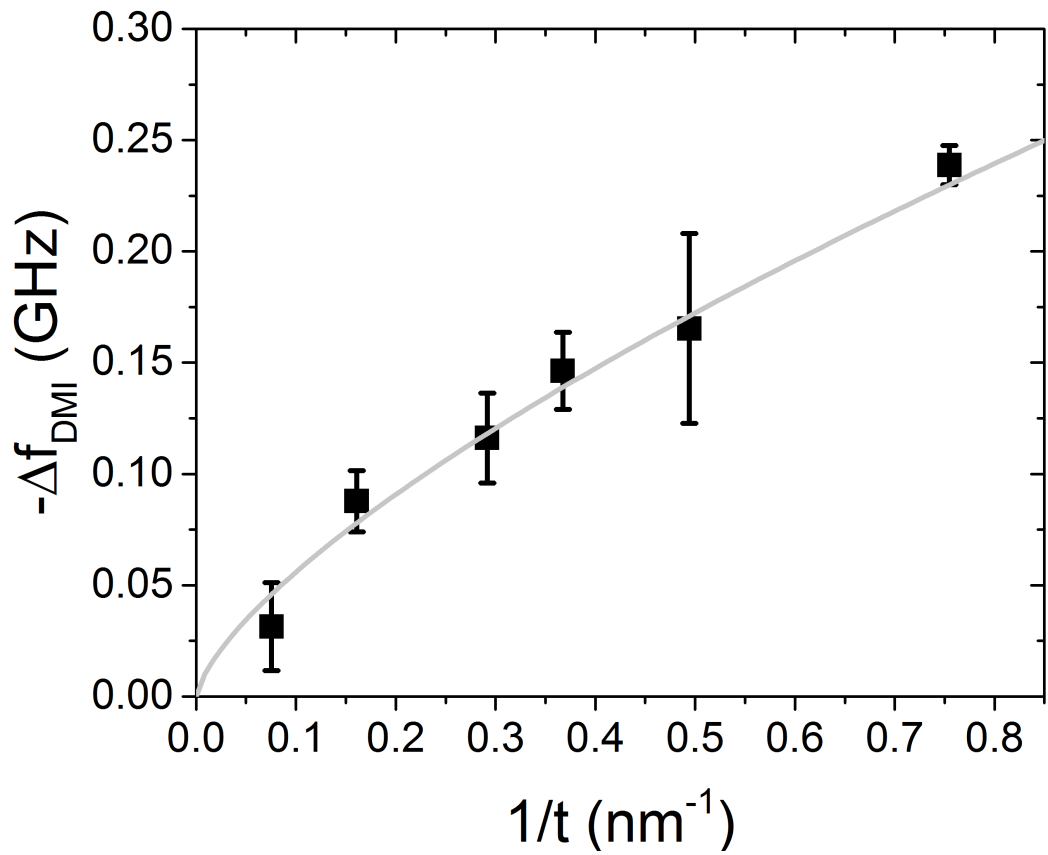

**Figure 3: DMI-induced spin-wave frequency shift.** Frequency shift $\Delta f_{DMI}$ as a function of the reciprocal $Ni_{80}Fe_{20}$ thickness. The grey line is a guide to the eye. $\Delta f_{DMI}$ increases with 1/t in agreement with the interfacial nature of the DMI.

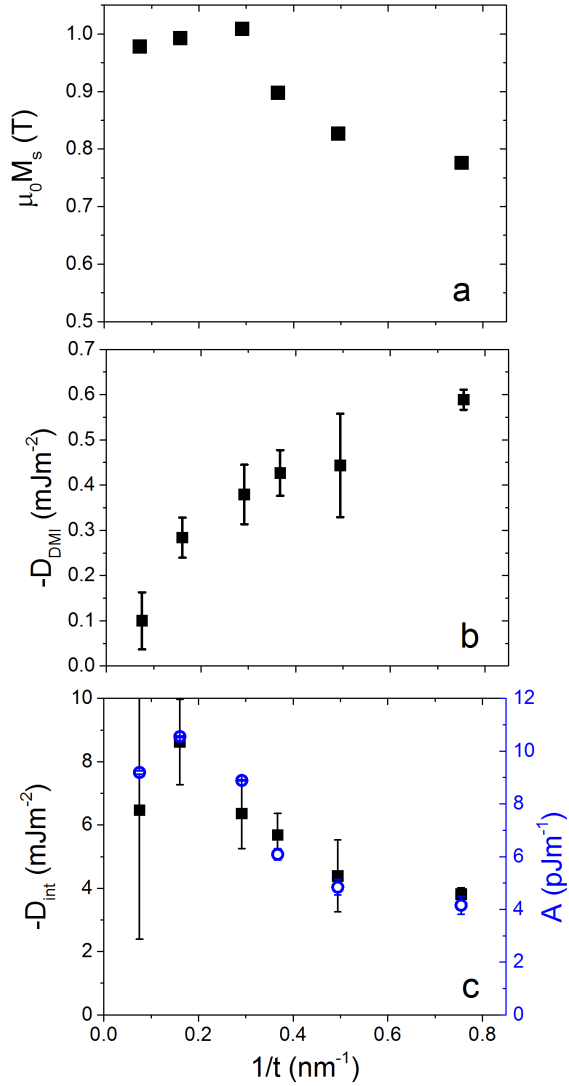

**Figure 4: Thickness dependence of the symmetric and the anti-symmetric exchange. a** The thickness dependence of the magnetization $M_s$ at 300 K. (The error bars are smaller than the symbol size.) **b** The volumetric DMI $D_{DMI}$ does not follow a strict $1/t$ dependence. **c** The antisymmetric exchange $D_{int}$ at the $Ni_{80}Fe_{20}$/Pt interface (squares, left scale) and the symmetric exchange $A$ in the $Ni_{80}Fe_{20}$ bulk (open circles, right scale) are linearly proportional over the entire range of measured thicknesses. The proportionality of $A$ and $D_{int}$ was originally predicted by Moriya[11]. (The error bars for $D_{DMI}$ and $D_{int}$ are the propagated errors of the averaged frequency shift. The error bars for $A$ reflect the variation of $A$ when using different exponents for the renormalization of exchange at non-zero temperatures, as explained in the SI.

## Determination of effective magnetic thickness for $Ni_{80}Fe_{20}$ films:

The physical thickness of the sputtered (111)-textured $Ni_{80}Fe_{20}$ layers was calibrated by use of x-ray reflectometry. The 0 K saturation magnetic moment $m_s$, obtained from the fit of $m_s$ vs. T SQUID magnetometer data to the Bloch $T^{3/2}$ law, is plotted against the physical thickness $t_n$ to determine the effective magnetic thickness of the $Ni_{80}Fe_{20}$ layer, as shown in Fig. S1. All samples were cut by a dicing saw into equal sized chips. The zero-moment intercept of a linear fit of the data indicates that there is a magnetic "dead" layer of thickness $t_D$=0.77 nm ±0.05 nm. The observation of such magnetic dead layers at the ferromagnet interface is not uncommon for sputtered metallic films[1]. Recently reported data from spin-pumping measurements provide evidence that the $Ni_{80}Fe_{20}$/Pt interface does not exhibit a dead layer[2] suggestive that the dead layer in the present case is at the $Ni_{80}Fe_{20}$/SiN interface, perhaps due to the formation of a non-magnetic silicide such as $Fe_nSi$ and/or $Ni_nSi$ during SiN deposition[3]. In the main text, we use the layer thickness $t=t_N-t_D$.

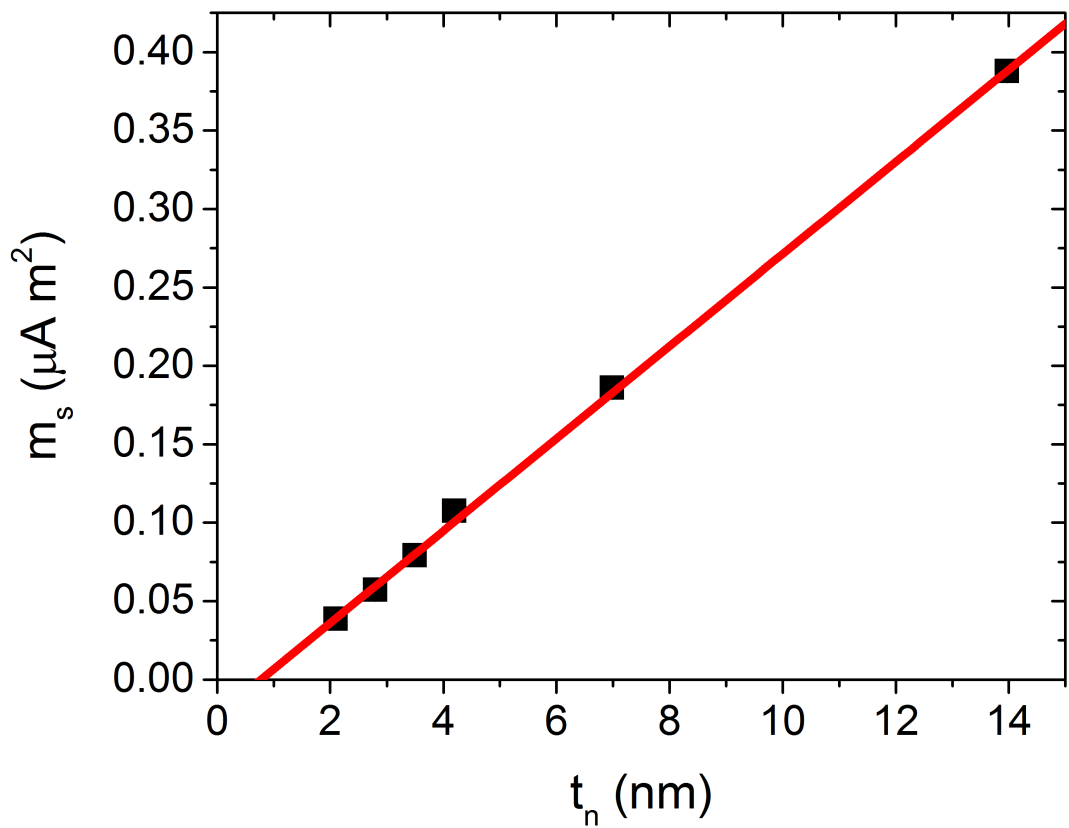

**Fig. S 1:** The magnetic moment is plotted versus the nominal thickness $t_n$ of the $Ni_{80}Fe_{20}$ layer. The non-zero x-axis intercept indicates the existence of a dead layer with a thickness $t_D$. The error is smaller than the symbol size.

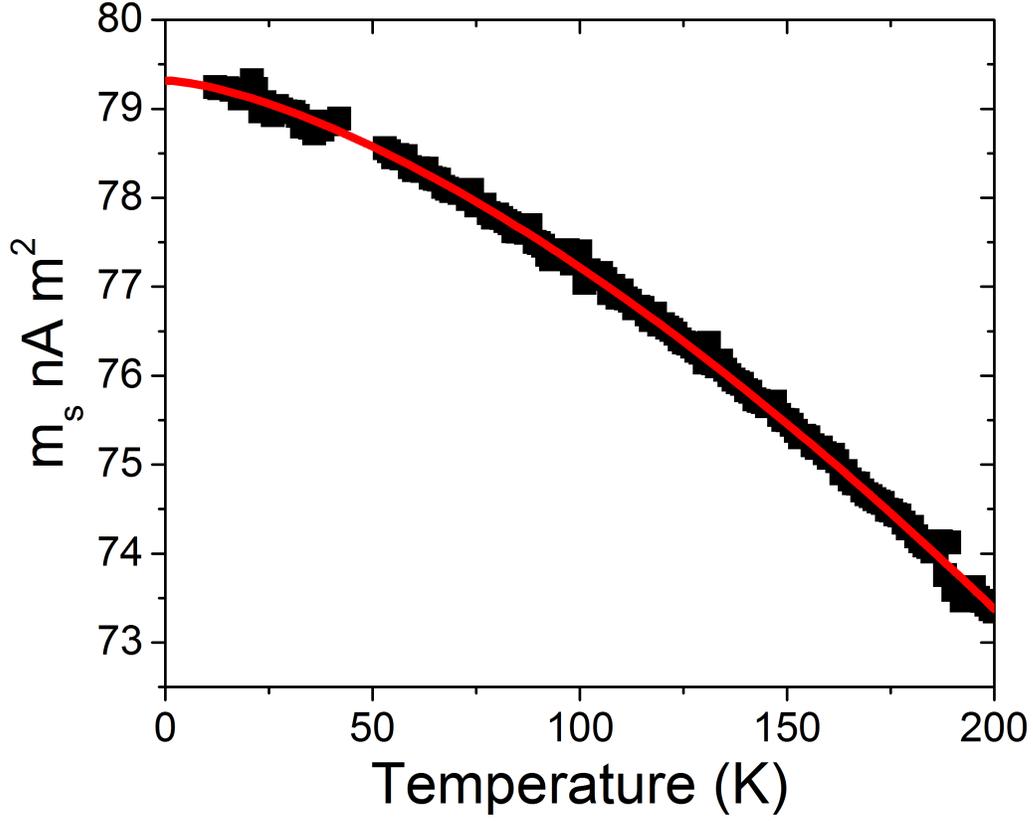

**Fig. S2**: Temperature-dependence of saturation moment $m_s$ for the $t=t_N\text{-}t_D=$ 2.7 nm sample. The red line is a fit of the data to the Bloch $T^{3/2}$ Law.

**Determination of the symmetric (Heisenberg) exchange constant:**

The Heisenberg Hamiltonian is

$$H = -\sum_{i \neq j} J_{ij}\left(\vec{S}_i \cdot \vec{S}_j\right), \quad \text{(S1)}$$

where $J_{ij}$ is the exchange energy between the two atomic sites $i$ and $j$, with spin $\vec{S}_i$ and $\vec{S}_j$, respectively. The symmetric-exchange contribution to the spin-wave energy $D_{\text{spin}}$ for a fcc lattice with lattice constant $a$ is given by

$$\begin{aligned} E_{\text{spin}} &= \frac{1}{3} S \sum_{i \neq j} J_{ij}\left(1 - \cos\left(\vec{k} \cdot \vec{r}_{ij}\right)\right) \\ &= \left(2 J_{nn} S a^2\right) k^2 \quad \text{(S2)} \\ &= D_{\text{spin}} k^2 \end{aligned}$$

where $\vec{r}_{ij}$ is the position vector connecting sites $i$ and $j$, $J_{nn}$ is the nearest-neighbor exchange integral, and $S$ is the total number of spins at each site.

In order to determine $D_{spin}$ we measured the temperature dependence of the magnetic moment $m_s(T,t)$ for Ni$_{80}$Fe$_{20}$ films of thickness $t$ for all samples by SQUID magnetometry. The data were fit with the Bloch T$^{3/2}$ Law:

$$m_s(T,t) = m_s(T=0\text{ K},t) \cdot \left( 1 - \frac{g\mu_B\eta}{M_s(T=0\text{ K},t)} \left( \frac{k_B T}{D_{spin}(T=0\text{ K},t)} \right)^{3/2} \right) \quad \text{(S3)}$$

where $k_B$ is Boltzmann's constant, $D_{spin}(T=0\text{ K},t)$ is the zero kelvin spin-wave stiffness with units of J m$^2$, and $\eta$ is a dimensionless constant that can depend on the sample dimensions. The prefactor $\frac{g\mu_B\eta}{M_s(T=0\text{ K},t)}$ equals the spin per unit cell. A plot of $m_s$ vs. $T$ for the t = 2.7 nm sample, together with the fit to Eq. (S3), is shown in Fig. S2. For a bulk sample, $\eta = 0.0587$. However, for thin films of reduced dimensionality used here, $\eta$ must be explicitly calculated. To calculate $\eta$, we used exact numerical summation of the Planck distribution for a finite volume with unpinned boundary conditions to determine the magnon density in thermal equilibrium. The detailed calculation of the magnon density is given in the section after the discussion of the anti-symmetric exchange.

In order to determine the spin-wave stiffness $D_{spin}(T=300\text{ K},t)$ at room temperature from $D_{spin}(T=0\text{ K},t)$, renormalization of the exchange needs to be taken into account. To lowest order the temperature dependence can be approximated with the temperature dependence of the saturation magnetization raised to the power of $\gamma$

$$D_{spin}(T,t) \cong D_{spin}(T=0\text{ K},t)\left[ \frac{M_s(T,t)}{M_s(T=0\text{ K},t)} \right]^{\gamma}. \quad \text{(S4)}$$

According to mean field theory[4] $\gamma = 1$, for an itinerant ferromagnet with magnon-electron interaction[5] $\gamma = 4/3$ and $\gamma = 0.7$ was found experimentally for magnetite[6]. In our manuscript we use $\gamma = 1$ from mean field theory and determine the error bars from the variation of $D_{spin}(T=0\text{ K},t)$ when using $\gamma = 4/3$ and $\gamma = 0.7$, respectively. (The particular choice of $\gamma$ does not alter the result, as indicated by the small size of the error bars for $A$ in the manuscript.) We then calculate the exchange constant $A$ at room temperature for each sample via

$$A(T=300\text{ }K,t) = D_{spin}(T=0\text{ K},t)\left[ \frac{M_s(T=300\text{ K},t)}{M_s(T=0\text{ K},t)} \right]^{\gamma} \cdot \left[ \frac{M_s(T=300\text{ K},t)}{2g\mu_B} \right] \quad \text{(S5)}$$

## Determination of the anti-symmetric (Dzyaloshinskii-Moriya Interaction) exchange

The DMI Hamiltonian for two atoms at sites *i* and *j* with the spins $\vec{S}_i$ and $\vec{S}_j$, respectively, is given by

$$H_{ij} = -\vec{D}_{ij} \cdot \left(\vec{S}_i \times \vec{S}_j\right) \quad \text{(S6)}$$

where $\vec{D}_{ij} \propto \vec{r}_{ij} \times \hat{n}$ is the Dzyaloshinskii-Moriya vector, with $\vec{r}_{ij}$ being the vector between sites *i* and *j* and $\hat{n}$ is a unit vector along the axis of broken symmetry, i.e. the interface normal, for our particular case. The following calculation applies to a single ferromagnetic monolayer in contact with a high spin-orbit material. Following the analysis of Udvardi and Szunyogh, for the case where a spin-wave propagates perpendicular to the magnetization, the DMI contribution to the spin-wave energy for an atomic monolayer is[7]

$$E_{DMI}^{int} = -cS \sum_{i \neq j} \left(\vec{D}_{ij} \cdot \hat{m}\right) \sin\left(\vec{k} \cdot \vec{r}_{ij}\right), \quad \text{(S7)}$$

where $\hat{m} \doteq \vec{M}/M_s$, $\vec{k}$ is the spin-wave wavevector, $S$ is the total spin on each site, and $c = \pm 1$ is the spin-wave chirality. (The necessary factor $S$ is missing in the original version of Eq. in Ref [7]). The chirality is defined by the orientation of the magnetic moment at each lattice site

$$\vec{e}_i(\vec{k},c) = \vec{n}_1 \cos\left(\vec{k} \cdot \vec{R}_i\right)\sin(\theta) + c\vec{n}_2 \sin\left(\vec{k} \cdot \vec{R}_i\right)\sin(\theta) + \hat{m}\cos(\theta).$$

$\vec{R}_i$ is the position vector for the atomic site *i*, $\theta$ is the relative angle of the moments and $\hat{m}$, $\vec{n}_1 \perp \hat{m}$ and $\vec{n}_2 = \vec{n}_1 \times \hat{m}$ are unit vectorsIn a (111)-textured fcc crystal the grains are randomly rotated around the surface normal and Eq. (S7) needs to be averaged about all possible orientations. Evaluation of Eq. (S7) in the long wavelength limit then yields

$$E_{DMI}^{int} \cong \frac{3cD_{nn}Sak}{\pi\sqrt{2}} \int_0^{2\pi} \cos(\theta)^2 \, d\theta$$
$$= \frac{3}{\sqrt{2}} cD_{nn}Sak \quad \text{(S8)}$$

where $\theta$ is the angle between $\vec{r}_{ij}$ and $\vec{k}$, $D_{nn}$ is the magnitude of the nearest neighbor DMI vector, and $a$ is the lattice constant.

## Calculation of the magnon density

For the determination of the spin-wave stiffness constant from the Bloch $T^{3/2}$ law, the prefactor $\eta$ in thermal equilibrium is required. We used exact numerical summation of the Planck distribution for a finite volume with unpinned boundary conditions to calculate. The magnon density at a temperature $T$ in thermal equilibrium is

$$n_m(T) = \frac{1}{8V} \sum_{i,j,k} \frac{1}{\exp\left(\frac{D_{spin} k_{i,j,k}^2 + hf_{FMR}}{k_B T}\right) - 1}, \quad (S9)$$

where

$$k_{i,j,k}^2 = \left(\frac{i\pi}{L_x}\right)^2 + \left(\frac{j\pi}{L_y}\right)^2 + \left(\frac{k\pi}{L_z}\right)^2. \quad (S10)$$

$L_{x,y,z}$ are the Cartesian dimensions of the finite-sized sample, $V = L_x L_y L_z$ is the volume of the magnetic sample, and $f_{FMR}$ is the FMR frequency in the field at which the temperature-dependent magnetization is measured. (As was the case for the original derivation of the $T^{3/2}$ law, only exchange-mode magnons are considered. Refinement of the calculation to include dipole field effects would greatly increase the complexity of the calculation, though it is not expected to have a larger impact on the final results, as is further explained below.) We transform Eq. (S9) into cylindrical coordinates and evaluate the sum numerically.

To determine $\eta$, we then divide the magnon density by the argument of the Bloch law, i.e.

$$\eta = \frac{n_m(T)}{\left(\frac{k_B T}{D_{spin}}\right)^{3/2}}. \quad (S11)$$

In Fig. S3, we plot of $\eta$ as a function of $t$ for the case of Permalloy with a lattice constant $a = 0.354$ nm. We see that $\eta$ is approximately equal to the classic bulk value of 0.0587 at the largest thickness calculated, but it deviates significantly from the bulk value for $t < 10$ nm.

The reason for the deviation can be easily understood in terms of how the Planck distribution is affected for spin waves with wavelengths comparable to or shorter than the film thickness. Spin waves with wavelengths less than 4 nm have an energy greater than $k_B T$. Referring to the Planck distribution, such modes are populated according to a Boltzmann-like factor of $\exp(-Dk^2/k_B T)$, such that they are only marginally occupied. However, in general, when the sample size is reduced, the reduction in the number of thermally excited magnons is balanced by the reduced volume of the sample, such that $n_m(T)$ is unaffected. However, the lowest order excitation of the solid, i.e. the FMR mode, is only weakly affected by the reduced sample volume, such that its occupation remains approximately constant. (In the spirit of the original derivation of the Bloch law, we are ignoring dipole field effects that would generally cause some modification of the lowest order excitation of the solid with reduced size. In particular, for the specific case considered here, the relative contribution of the exchange energy to the mode frequency, i.e. > 10 meV, greatly exceeds the contribution due to dipole fields, which is more along the lines of only 0.1 meV.) Thus, once all the spin-wave modes have been frozen out due to a reduced dimension of less than 4 nm along a particular quantization axis, the FMR mode continues to be highly occupied in spite of any further reductions in the sample size. This results in an enhancement of $n_m(T)$, and therefore $\eta$, with continued decrease in the sample thickness.

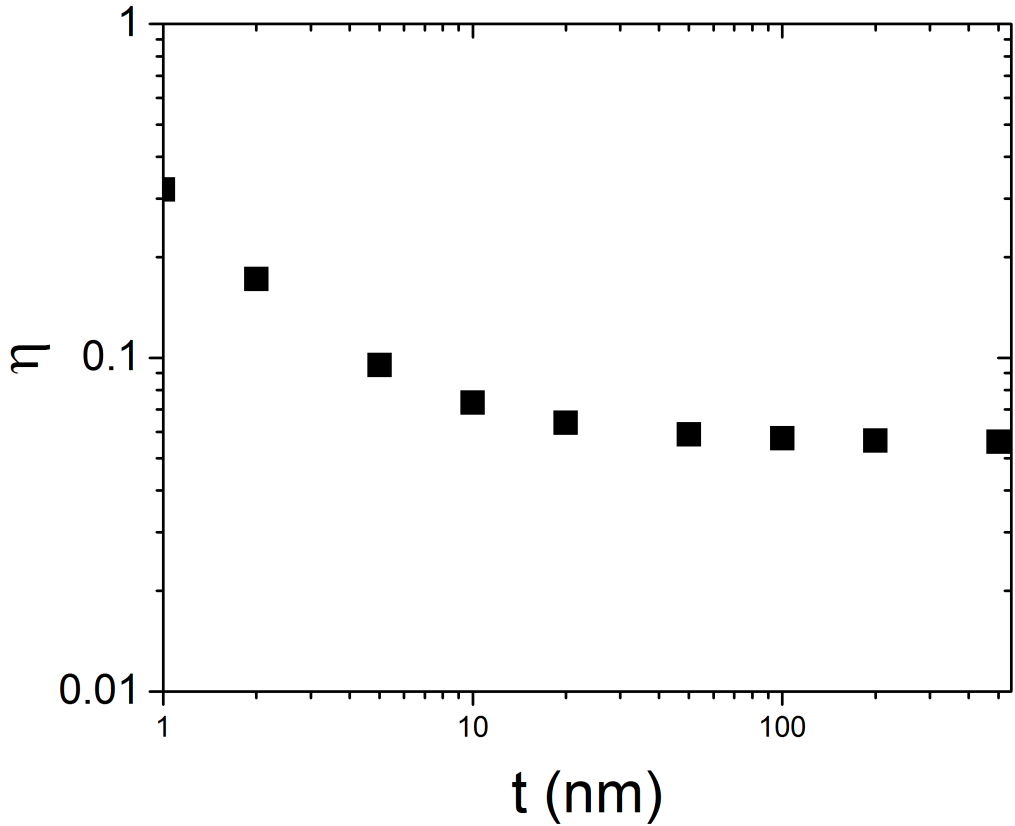

**Fig. S3:** Prefactor $\eta$ for the Bloch $T^{3/2}$ law. For thicknesses $t > 10$ nm $\eta$ approaches the bulk value 0.0587.

**Ferromagnetic Resonance Measurements**

We determined the net perpendicular anisotropy and in-plane spectroscopic $g$-factor $g^P$, see Fig. S4 and S5, by use of broadband vector network analyzer ferromagnetic resonance (VNA-FMR) measurements for the cases of a saturating magnetic field both perpendicular-to and parallel-to the sample plane. Experimental details can be found in Nembach et al[8]. The data for the resonance field $H_\perp$ vs. excitation frequency $f$ data in the out-of-plane geometry are fitted to the Kittel equation for the perpendicular geometry,

$$f = \frac{\mu_0 \mu_B g^\perp}{h}\left(H_\perp - M_{eff}\right), \text{ (S12)}$$

where $\mu_0$ is the permeability of free space, $\mu_B$ is the Bohr magneton, $h$ is Planck's constant, $g^\perp$ is the spectroscopic splitting factor for the perpendicular geometry, and $M_{eff} = M_s - H_k$ is the effective magnetization, where $H_k$ is the interfacial perpendicular anisotropy field. For $H_k > 0$, the interfacial anisotropy easy axis is parallel to the sample normal. The fitting parameters are $g^\perp$ and $M_{eff}$. Together with the room temperature values for $M_s(T = 300\text{ K}, t_n)$ obtained from SQUID magnetometry, we determine the thickness-dependence of $H_k$, shown in Fig. S4. The linear dependence of $H_k$ on $1/t$ is a clear signature of interfacial anisotropy.

The in-plane VNA-FMR resonance fields $H_\parallel$ are also fitted with the Kittel equation for this geometry.

$$f = \frac{\mu_0 \mu_B g^\parallel}{h}\sqrt{H_\parallel \cdot (H_\parallel + M_{eff})}. \text{ (S13)}$$

Since we have previously shown that the perpendicular and in-plane values of $M_{eff}$ are identical[9], $g^\parallel$ shown in Fig. S5 is more precisely determined from Eq. (S12) by fixing the value of $M_{eff}$ to that obtained in the perpendicular geometry.

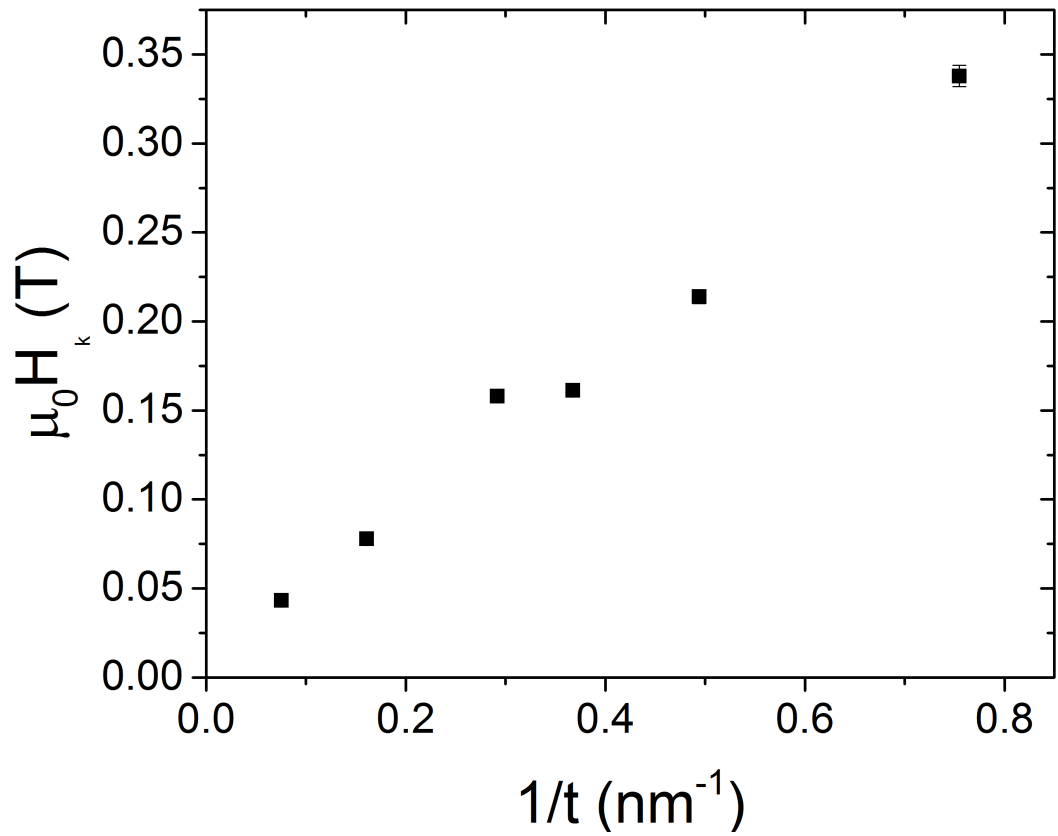

**Fig. S4:** Anisotropy field $H_k$ versus the reciprocal thickness $t$. The error is smaller than the symbol size and originated from the fit of the data to the Kittel equation.

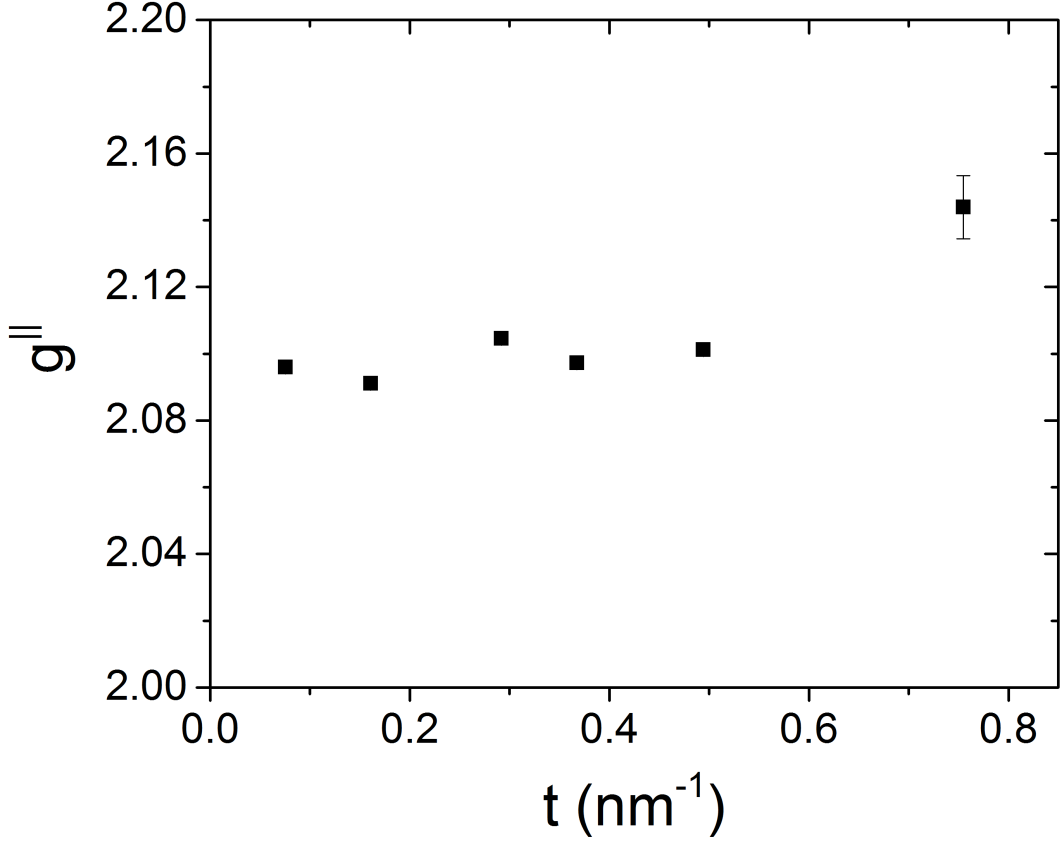

**Fig. S5:** In-plane spectroscopic splitting factor $g^{\parallel}$ versus the reciprocal thickness $t$. The error originated from the fit of the data to the Kittel equation.

### Damon-Eshbach Modes with anti-symmetric exchange

Following the analysis in Moon et al.[10] the spin-wave dispersion for spin-waves in a thin film with $\vec{k} \perp \vec{M}$ for in-plane magnetization, i.e. the Damon-Eshbach mode (DE), in the presence of interfacial DMI is[10] [11] [12]

$$\omega = \frac{\mu_0 \mu_B g^{\parallel}}{\hbar} \sqrt{\left(H + \frac{2A}{\mu_0 M_s}k^2 - H_K + M_s \frac{1-e^{-t|k|}}{t|k|}\right) \cdot \left(H + \frac{2A}{\mu_0 M_s}k^2 + M_s\left(1 - \frac{1-e^{-t|k|}}{t|k|}\right)\right)}$$

$$+ \text{sgn}(H) \frac{\mu_B g^{\parallel}}{\hbar} \cdot \frac{2D_{DMI}}{M_s} k \quad \text{(S14)}$$

where $A$ is the exchange constant. The last term in Eq. (S14) accounts for the DMI, which causes a frequency shift linear in the magnitude of the spin-wave wavevector. The sign of the DMI-induced frequency shift depends on the sign of the magnetization and the propagation direction of the spin-waves. Comparison with Eq. (S14), and accounting for the conversion from the interfacial DMI to the volumetric DMI, as quantified in Eq. 3 in the manuscript, allows us to express the volumetric DMI constant $D_{DMI}$ in terms of the microscopic $D_{nn}$ as:

$$D_{DMI} = \frac{3}{2\sqrt{2}}\left(\frac{a}{\sqrt{3}t}\right)\frac{M_s}{\mu_B g^{\parallel}}\left(cD_{nn}Sa\right) \quad \text{(S15)}$$

In terms of the interfacial DMI constant $D_{DMI}^{int}$ we have

$$D_{DMI}^{int} = \frac{3}{2\sqrt{2}}\frac{M_s}{g\mu_B}\left(cD_{nn}Sa\right) \quad \text{(S16)}$$

In the main part of the manuscript we calculate the exchange constant $A$

$$A = \frac{M_s a^2 S J_{nn}}{g\mu_B}. \quad \text{(S17)}$$

The ratio of the interfacial DMI constant and the exchange constant can now be related to the fundamental quantities of the symmetric and antisymmetric exchange:

$$\frac{D_{DMI}^{int}}{A} = \frac{3}{2\sqrt{2}}\frac{1}{a}\left(\frac{D_{nn}}{J_{nn}}\right) \quad \text{(S18)}$$

Thus, if the asymmetric exchange simply scales in proportion to the symmetric exchange for a given material system, the ratio of the interfacial DMI and the exchange constant should be constant.

For thin magnetic films, the DE spin-wave modes are non-reciprocal, i.e. the mode is localized either at the top or bottom interface, depending on the propagation direction relative to $\vec{M}$. If the magnetization direction is reversed, the spin-waves are localized at the opposite interface. Thus, if the two Ni$_{80}$Fe$_{20}$ interfaces were to have strongly differing magnetic properties, it is possible that the non-reciprocal nature of the DE modes could also give rise to an asymmetric spin-wave dispersion. In the following, we will estimate an upper value for the frequency shift due to the non-reciprocal character of the surface-waves.

For the geometry used here, the spin-wave propagating positive x-direction is localized at the Pt interface for positive applied field, whereas the spin-wave propagating in the opposite direction is localized at the SiN interface. If we neglect the symmetric-exchange contribution, i.e. the long wavelength limit, the DE-mode amplitude decays exponentially into the film thickness with a decay length $\delta = 1/k$. For the wavevector of the spin-waves, which are measured in the BLS experiment, this results in a decay length is $\delta$= 60 nm. This results in a difference of the spin-wave amplitude at both interfaces by 2% and 20% for the thinnest and thickest sample of the $Ni_{80}Fe_{20}$/Pt series, respectively. The Pt interface induces a perpendicular anisotropy field $H_k$, which thickness dependence is shown in Fig. S4. Thus the frequency of the spin-wave, which is stronger localized at the Pt interface, is shifted to lower frequencies with respect to the spin-wave propagating in the opposite direction. In order to calculate the frequency shift due to the surface anisotropy we use a mean-field approach. The anisotropy field $H_k^{\text{int}}(t)$ is assumed to be localized within the first monolayer of the $Ni_{80}Fe_{20}$ layer. The effective anisotropy for each spin-wave propagating in positive and negative direction, respectively, is then calculated by weighing the anisotropy field with the spin-wave amplitude for the respective mode profile $m^{\pm}(t)$

$$H_k^{\text{eff}\pm} = \frac{\int H_k^{\text{int}}(t) m^{\pm}(t) dt}{\int m^{\pm}(t) dt} \quad , \text{(S19)}$$

The integrals are extending over the $Ni_{80}Fe_{20}$ layer thickness. The estimate of the frequency difference for the spin-waves propagating in opposite directions is then calculated with Eq. (S14). We find a frequency shift for the samples of about 0.03 GHz almost independent of the sample thickness. This frequency shift would be even smaller if exchange would be taken into account, because it would lead to a more gradual decrease of the spin-wave amplitude from the respective surface. Hence the value of 0.03 GHz is an upper limit for the frequency shift due to the localization of the Damon-Eshbach modes. Consequently, the frequency shift due to the non-reciprocal character of the surface spin-waves has only a very marginal influence on the frequency-shifts determined by BLS and is therefore neglected.

1. Lefakis, H. *et al.* Structure and magnetism of Ta/Co/Ta sandwiches. *J. Magn. Magn. Mater.* **154,** 17–23 (1996).

2. Boone, C., Shaw, J. M., Nembach, H. T. & Silva, T. J. *ArXiv14085921v1 (2014)*

3. Schlesinger, M. E. Thermodynamics of solid transition-metal silicides. *Chem. Rev.* **90,** 607–628 (1990).

4. Atxitia, U. *et al.* Micromagnetic modeling of laser-induced magnetization dynamics using the Landau-Lifshitz-Bloch equation. *Appl. Phys. Lett.* **91,** 232507 (2007).

5. Lovesey, S. W. *Theory of neutron scattering from condensed matter*. (Clarendon Press, 1984).

6. Heider, F. & Williams, W. Note on temperature dependence of exchange constant in magnetite. *Geophys. Res. Lett.* **15,** 184–187 (1988).

7. Udvardi, L. & Szunyogh, L. Chiral Asymmetry of the Spin-Wave Spectra in Ultrathin Magnetic Films. *Phys. Rev. Lett.* **102,** (2009).

8. Nembach, H. T. *et al.* Perpendicular ferromagnetic resonance measurements of damping and Landég– factor in sputtered $(Co_2Mn)_{1-x}Ge_x$ thin films. *Phys. Rev. B* **84,** 054424 (2011).

9. Shaw, J. M., Nembach, H. T., Silva, T. J. & Boone, C. T. Precise determination of the spectroscopic g-factor by use of broadband ferromagnetic resonance spectroscopya). *J. Appl. Phys.* **114,** 243906 (2013).

10. Moon, J.-H. *et al.* Spin-wave propagation in the presence of interfacial Dzyaloshinskii-Moriya interaction. *Phys. Rev. B* **88,** 184404 (2013).

11. Demokritov, S. & Tsymbal, E. Light scattering from spin waves in thin films and layered systems. *J. Phys. Condens. Matter* **6,** 7145 (1994).

12. Stamps, R. L. & Hillebrands, B. Dipolar interactions and the magnetic behavior of two-dimensional ferromagnetic systems. *Phys. Rev. B* **44,** 12417 (1991).